\documentclass{pasj00}
\draft
\SetRunningHead{Astronomical Society of Japan}{Usage of \texttt{pasj00.cls}}

\usepackage{times}

\begin{document}

\title{Spatial and Temporal Variations of the Diffuse Iron 6.4 keV Line
in the Galactic Center Region}

\author{Dmitrii \textsc{Chernyshov}$^{1,2,3}$, Vladimir
\textsc{Dogiel}$^{1}$}
 \affil{$^1$I.E.Tamm Theoretical Physics Division of P.N.Lebedev  Institute of Physics,
Leninskii pr. 53, 119991, Moscow, Russia} \affil{$^2$Department of
Physics, The University of Hong Kong, Pokfulam Road, Hong Kong,
China} \affil{$^3$Institute of Astronomy, National Central
University, Jhongli 320, Taiwan} \email{chernyshov@td.lpi.ru}
\author {Masayoshi \textsc
{Nobukawa}, Takeshi \textsc{Go Tsuru}, Katsuji \textsc{Koyama}}
 \affil{Department of Physics, Graduate
school of Science, Kyoto University, Oiwake-cho, Kitashirakawa,
Kyoto 606-8502}
\author{Hideki \textsc {Uchiyama}}
\affil{Department of Physics, School of Science, The University of
Tokyo, 7-3-1 Hongo, Bunkyo-ku, Tokyo 113-0033} \and
\author{Hironori \textsc {Matsumoto}}
\affil{Kobayashi-Maskawa Institute for the Origin of Particles and
the Universe, Nagoya University, \\
Furo-cho, Chikusa-ku, Nagoya,
464-8602}

 \KeyWords{Galaxy: center --- X-rays: ISM  --- cosmic-rays}

\maketitle

\begin{abstract}
We analyze the diffuse  Fe \emissiontype{I} K $\alpha$ line
generated in the diffuse interstellar molecular hydrogen by
primary photons or subrelativistic protons injected by Sagittarius
(Sgr) A$^\ast$. We showed that unlike emission from compact
molecular clouds, this emission can be permanently observed in the
directions of the Galactic center. We conclude that the diffuse
emission of 6.4 keV line observed at present is probably due to 
Fe\emissiontype{I} K $\alpha$ vacancy production by primary photons if the X-ray
 luminosity of Sgr A$^\ast$ was about $L_X \sim 10^{39}-10^{40}$ erg
s$^{-1}$. In principle these data can also be described in the
framework of the model when the 6.4 keV line emission is generated
by subrelativistic protons generated by accretion onto the central
black hole but in this case extreme parameters of this model are
necessary.

\end{abstract}

\section{Introduction}
Detection of the Fe\emissiontype{I} K$\alpha$ (6.4 keV) line from
molecular clouds in the Galactic center (GC) is one of the most
remarkable events in the high energy astrophysics of the last
decades. The story had started from 1993 when \citet{suny} found a
diffuse X-ray emission from compact sources in the GC. They
interpreted this emission as due to reflection of photons by dense
molecular clouds (Compton echo) which were irradiated by a nearby
X-ray source. In addition, they predicted a bright fluorescent
K$\alpha$ line in the scattered spectrum of the clouds due to the
K-absorption of photons with energies $E>7.1$ keV. This line was
indeed discovered soon afterwards by \citet{koya1}. The brightest
region of the 6.4 keV emission was located over the giant
molecular cloud Sgr B2. Later on the 6.4 keV emission was
discovered also in other molecular clouds of the GC (see
\cite{mura2,nobuk,bamba}).
\subsection{X-ray Reflection Nebula (XRN) model}
 \citet{koya1}  (see also
\cite{mura00}) speculated that the 6.4 keV flux from Sgr B2 is due
to the past activity of the the Galactic nucleus Sgr A$^\ast$
which had been bright several hundred years ago but is currently
dim. From the observed flux of 6.4 keV photons from the cloud,
$F_{6.4}\sim 10^{-4}$ ph cm$^{-2}$ s$^{-1}$, and the gas column
density of the cloud the differential spectrum of primary photons
$dn(E_x)/dE_x$ can be estimated from the equation (see e.g.
\cite{tatis})
\begin{equation}
F_{6.4}\sim\frac{c}{4\pi R_\odot^2}\int\limits_{V_{Sgr B2}}d^3r
\int\limits_{E_x>7.1~keV}
n_H(r)\frac{dn(E_x,r)}{dE_x}{dE_x}\varepsilon (E_x,r)dE_x
\label{est}
\end{equation}
where $V_{Sgr B2}$ is the volume of the cloud, $\varepsilon
(E_x,r)$ is the emissivity of 6.4 keV photons at the coordinate
$r$ inside the cloud and $R_\odot$ is the distance to Earth. The
spectrum of primary photons can be derived from the observed
continuum emission from Sgr B2 which in the case of the XRN
scenario is the same as the spectrum of primary photons.
\citet{mura00} showed that these spectrum
 was a power-law with the spectral index of $-2$ in the energy range 2--10
 keV (see also \cite{koya09})
\begin{equation}
\frac{dn(E_x)}{dE_x}\propto E_x^{-2}\,.
 \label{koya_sp}
 \end{equation}
   The 2--10 keV luminosity of primary photons
from Sgr A$^\ast$  necessary to produce the observed 6.4 keV
emission from Sgr B2 was estimated  by \citet{mura00},
\begin{equation}
L_{2-10~keV}\sim 3\times
10^{39}\left(\frac{d}{100~pc}\right)^2\mbox{erg s$^{-1}$}
\label{f_sgr}
\end{equation}
where $d$ is the distance between Sgr A$^\ast$ and Sgr B2.

The most direct evidence to favor the photoionization  by an X-ray
flash of the central source would be a time variability of the 6.4
keV line emission  from molecular clouds because of a relatively
short time in which a photon crosses them. This idea about the Sgr
A${^\ast}$ past activity as the origin of 6.4 keV emission from
Sgr B2 was confirmed recently. Observations found a steady
decrease of the X-ray flux from Sgr~B2 for the period $\lesssim
10$ years. Time variations of the emission are expected in the XRN
model and interpreted as photoionization of iron atoms by a flux
of primary X-rays  emitted by the central source Sgr A$^\ast$ due
to an X-ray flare occurred there about 100--300 years ago
\citep{koya08,inui,terrier,nobukawa1}.

 \subsection{Alternative Models of 6.4 keV Line Emission}
 In principle a flux of the  6.4 keV line can also be generated by collisions of
 subrelativistic charged particles with the molecular gas in the
 GC. Thus, \citet{yus1}  accounted the impact of
subrelativistic electrons with energies 10 -- 100 keV from local
sources with diffuse neutral gas producing both nonthermal
bremsstrahlung X-ray continuum emission and diffuse 6.4 keV line
emission. \citet{dog0} suggested a  scenario for the 6.4 keV line
emission from molecular clouds  which was excited by a flux of
subrelativistic protons produced by star accretion onto the
central black hole. Below these two scenarios are denoted as the
low energy cosmic-ray electron model (LECRe) and the low energy
cosmic-ray proton model  (LECRp). We note that because of
relatively long life time of protons in comparison  with the
average time of star capture the LECRp component of 6.4 keV line
emission in the GC is quasi-stationary.

Generation of the 6.4 keV line is accompanied by a continuum X-ray
emission produced by the Thomson scattering for the XRN scenario
and by bremsstrahlung for the LECRe and LECRp models. Therefore
the origin of 6.4 keV line flux from the clouds  can be defined
from the analysis of the equivalent width {\it eW} of the iron
line in the spectrum  which is the ratio
 of the line flux to the continuum intensity at $E_x=6.4$ keV,
 \begin{equation}
eW=\frac{F_{6.4}}{F_x(E_x=6.4~keV)}\,.
\end{equation}
The width  is a function of the iron abundance $\eta$ in the
clouds.

From  estimations of {\it eW} for the cloud Sgr B2
\citet{nobukawa} concluded  that the photoionization
interpretation seemed to be more attractive in comparison with the
electron impact scenario. The required abundance of iron in the
cloud was estimated by \citet{nobukawa1} by the value $\eta=1.3$
solar.
However, \citet{capelli}  might find  the iron line emission which
was produced by subrelativistic particles.  They presented results
of eight years of XMM-Newton observations of the region
surrounding the Arches cluster in the Galactic Center. They
analyzed spatial distribution and temporal behavior of the 6.4 keV
line emission and concluded that the origin of this emission might
be of the photoionization origin, although excitation by
cosmic-ray particles was not excluded. Moreover, they concluded
that for the three clouds nearest to the Arches cluster, which
showed a constant flux over the 8-year observation, the origin of
the line as photoionization by photons from Sgr A$^\ast$ seemed to
be at best tentative, and the hardness of the nonthermal component
associated with the 6.4-keV line emission might be best explained
in terms of bombardment by cosmic-ray particles.

Recent Suzaku observations might  also find the iron line emission
which was produced by subrelativistic particles \citep{fuku,
tsuru}. For the clumps G\,0.174$-$0.233 with $eW\simeq 950$ eV
they concluded that the  XRN scenario was favored. On the other
hand, for the clump G\,0.162$-$0.217 with $eW\simeq 200$ eV they
assumed that the emission from there was due to  low energy
cosmic-ray electron (LECRe). They found also that the $eW$ of the
6.4 keV emission line detected in the X-ray faint region (non
galactic molecular cloud region)  was significantly lower  than
one expected in the XRN scenario but higher than that of the LECRe
model.

 \citet{dog11} showed that estimates of  {\it eW} alone did not allow to
 distinguish firmly between the XRN and LECR scenarios because in the latter case the
 value of {\it eW} depended strongly on a spectral index of ionizing charged particles,
 especially if they were subrelativistic protons.
 In the case of ionization by charged
 particles spatial characteristics of 6.4 keV line are expected
 quite different for electrons and protons. While for
electrons we expect rather local ionization of the medium because
of their  relatively short lifetime, protons can fill an extended
region around the GC.
If protons generate 6.4 keV line in the GC then  at least two
components of  6.4 keV line emission from the molecular clouds and
the diffuse molecular gas can be generated there. The first one is
a time variable component generated by a flare of primary X-rays
emitted by Sgr A$^\ast$, and the second is a quasi-stationary
component produced by subrelativistic protons interacting with the
gas.

Observations of the 6.4~keV flux from Sgr B2 have not found up to
now any evident stationary component for the GC molecular clouds,
though as predicted by \citet{ponti} a fast decrease of 6.4 keV
emission observed with XMM-Newton for several molecular clouds
suggested that the emission generated by low energy cosmic-rays,
if present, might become dominant in several years.
A component of another than that of the XRN origin may also be
seen in the X-ray spectrum of  faint molecular regions in the GC
as follows from \citet{fuku}. Below we  derive parameters of the
diffuse  6.4 keV line emission in the framework of the XRN and
LECRp models in attempts to define the origin of the observed
diffuse line flux from the GC.

\section{Diffuse Emission of the 6.4 keV Line   from the GC}
The intensity of the diffuse 6.4 keV line
 from the GC depends on  parameters of the intercloud molecular gas there.
 The inner bulge (200--300 pc
central region) contains $(7-9)\times 10^7~M_\odot$ of hydrogen
gas. In spite of relatively small radius this region contains
about 10\% of the Galaxy's molecular mass. Half of the molecular
gas  is contained in very compact clouds of mass
$10^4-10^6M_\odot$, average densities of which is $\geq 10^4$
cm$^{-3}$  with a volume filling factor of only a few per cent,
then the cloud radius is in the range 1--40 pc. The other half
forms the molecular intercloud gas with the densities of at least
$n_H>10-10^2$ cm$^{-3}$ (see \cite{morris}). \citet{laun02}
estimated for the inner $\sim 200$ pc region the averaged
molecular hydrogen density $n_H$ to be  140 cm$^{-3}$ assuming
homogeneous matter distribution, and for a thin intercloud medium
$n_H\sim 10$ cm$^{-3}$.

Recently \citet{koya09} provided careful analysis with high energy
resolution and low background of the diffuse 6.4 keV emission and
of the hard X-ray continuum  associated with this line in the GC
region. From the Suzaku data they estimated the continuum X-ray
emission which is proportional to the intensity of diffuse 6.4 keV
line.

\begin{figure}[ht]
\begin{center}
\FigureFile(130mm,130mm){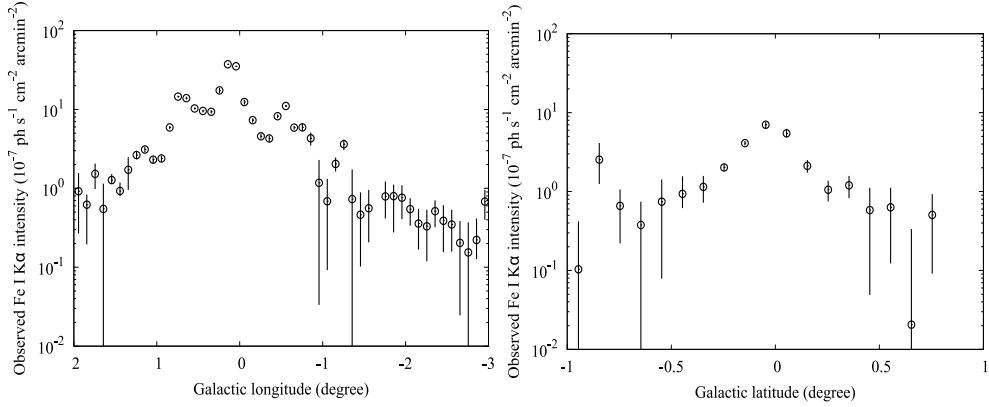}
\end{center}
\caption{Longitude (along $b=\timeform{-0D.046}$)  and latitude
($l=\timeform{-0D.17}$) distributions of the 6.4 keV line in the
GC as observed by Suzaku. The data-points taken from
\citet{uchi11}}\label{6.4_dist_l}
\end{figure}

 More recently \citet{uchi11} provided a careful
analysis of 6.4 keV emission from the region around the GC.  
Their spatial
 distribution of 6.4 keV line in the GC is shown in figure \ref{6.4_dist_l}
 where spikes of this emission
 correspond to directions to molecular clouds.

We expect that characteristics of the 6.4 keV emission produced by
subrelativistic cosmic-rays and by a flux of primary X-ray photons
are quite different. Below we reproduce  spatial distributions of
the 6.4 keV line in the framework of  XRN and LECRp models.

\section{Spatial Distribution of the Diffuse 6.4 keV Line Emission in the XRN
Model}

It is assumed in the XRN model  that  Sgr A$^\ast$ was active for about $T_1$ years
in the past as an emitter of primary X-ray photons with the energy  $E_x$. The average
luminosity of this source during the active period is
\begin{equation}
L_{fl}\simeq 10^{39} \mbox{erg s$^{-1}$}
\end{equation}
for the range 2--10 keV. The source activity is supposed to
cease $T$ years ago.

For the observed spectrum (\ref{koya_sp}) we define the total density of
primary photons on the divergent front of primary photons, which is at the
distance $r$ from Sgr A$^\ast$, as
\begin{equation}
n_{ph} = \frac{L_{fl}}{4\pi cr^2 E_x^{min}
\ln(E_x^{max}/E_x^{min})} \label{nph}
\end{equation}
where $E_x^{min}$ and $E_x^{max}$ are the minimum and maximum
energies of the spectrum of primary photons. Then the differential
spectrum of primary photons $dn(E_x)/dE_x$ for the total photon
density (\ref{nph}) we present as
\begin{equation}
\frac{dn(E_x)}{dE_x}=n_{ph}F_x(E_x)
\end{equation}
whith the normalization condition for $F_x(E_x)$
\begin{equation}
\int\limits_{E_x^{min}}^{E_x^{max}}F_x(E_x)dE_x=1
\end{equation}
Therefore for the spectrum (\ref{koya_sp}) we have
\begin{eqnarray}
&&F_x(E_x)=\frac{E_x^{min}E_x^{max}}{(E_x^{max}-E_x^{min})
}E_x^{-2}\theta(E_x-E_{min})\theta(E_{max}-E_x)\simeq\nonumber\\
&&\simeq E_x^{min}
E_x^{-2}\theta(E_x-E_{min})\theta(E_{max}-E_x)\,,
~~~\mbox{for $E_{max}>>E_{min}$}
\end{eqnarray}
Here $\theta(y)$ is the Heaviside step function.

These primary photons ionize iron atoms. The cross-section of
photoionization $\sigma_K$ has a form
\begin{equation}
\sigma_K(E_x) = \sigma_0
\left(\frac{E_x}{E_0}\right)^{-3}\theta(E_x-E_0)
\end{equation}
 where $E_0 = 7.1$
keV and $\sigma_0 \sim 3\times 10^{-20}$ cm$^2$ (see
\cite{tatis}).

Then the emissivity of the 6.4 keV line is
\begin{eqnarray}
\epsilon_{6.4}(r) =&& c\eta\omega_K n_H
n_{ph}\int\limits_{E_0}^{E_{max}} \sigma_x(E_x) F_x(E_x) dE_x =
\nonumber \\
=&& c\eta\omega_K\sigma_0 n_H n_{ph} \frac{E_x^{min}}{4E_0}\left(1-\frac{E^4_0}{E^4_{max}}\right) \simeq
\eta\omega_K\sigma_0 n_H \frac{L_{fl}}{4\pi r^2
 \ln(E_x^{max}/E_x^{min})} \frac{1}{4E_0}\,,
\end{eqnarray}
where $\omega_K$ is the fluorescence yield of X-ray photon
emission, which is about 0.3 for iron. The average density of the
diffuse molecular gas was defined as $n_H$. Below we take
everywhere for calculations $n_H=10$ cm$^{-3}$ and assume a
uniform density distribution of the molecular gas in the GC that
gives an upper limit of diffuse 6.4 keV emission from the GC. The
iron abundance $\eta$ is supposed to equal twice solar,
$\eta=2\eta_\odot\simeq 7\times 10^{-5}$.

\begin{figure}[ht]
\begin{center}
\FigureFile(100mm,100mm){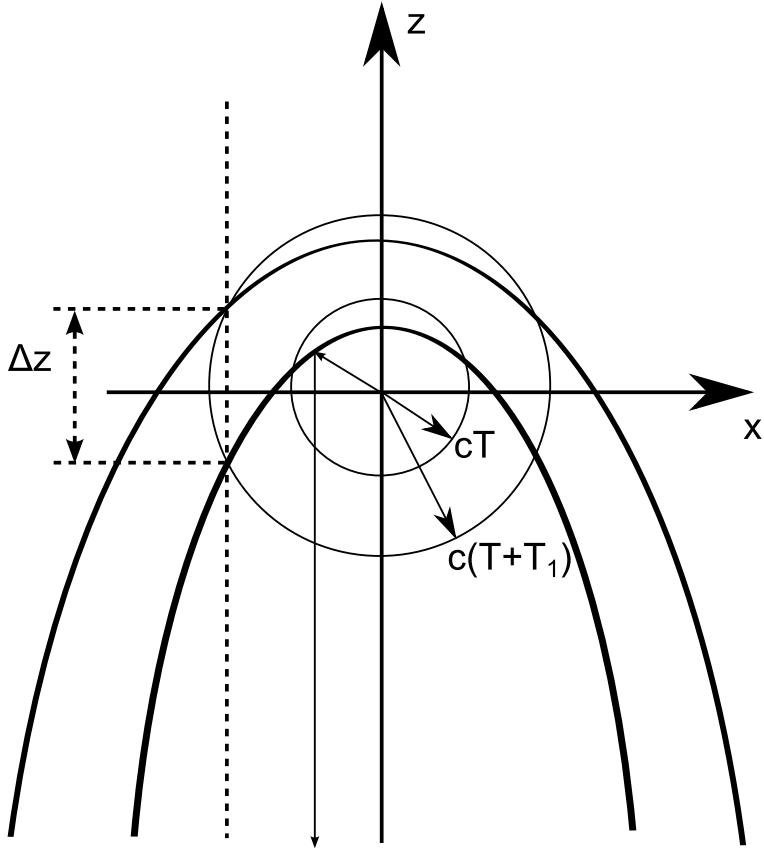}
\end{center}
\caption{Geometry of the GC reflection process. The source Sgr
A$^\ast$ in the coordinate center. Two parables shown by solid
lines denote the reflection positions of emission emitted in the
time interval $\{\tau_1, \tau_2\}$ which can be observed by an observer
at present. Two circles (thin  lines) denote a schematic position of the
front of primary X-ray photons emitted by  Sgr A$^\ast$ for the period $T_1$ which stopped its activity   $T$ years ago. The
dashed line is the line of view of the
observer. Two thin arrow lines show the path of a primary photon before and after reflection.}
\label{geom}
\end{figure}

For the delay time $T$ we can observe at present an irradiate emission
of the diffuse gas which is on surface of the parabola (see e.g.
\cite{suny1})
\begin{equation}
\frac{z}{c}=\frac{1}{2T}\left[T^2-\left(\frac{x}{c}\right)^2\right]\,,
\label{surf}
\end{equation}
where the coordinates $x$ and $z$ are  shown in figure \ref{geom}.

Unlike the line emission from compact molecular clouds which can
be observed for a relatively short period of time  when the X-ray
front is crossing  a cloud ($\sim 10$ years), the diffuse 6.4 keV
emission produced by primary X-ray photons should be permanently
observed from the GC as the front of primary X-ray photons
is propagating though the diffuse molecular gas in the GC.

The region emitting the 6.4 keV line by the diffuse gas -- X-ray
photon interactions is enclosed between the two surfaces
determined by the time $\tau_1=T$ and $\tau_2=T+T_1$ (see Eq.
(\ref{surf})) whose thickness  is $\Delta z$ between $z_1$
corresponding $t=\tau_1$ and $z_2$ corresponding $t=\tau_2$. We
showed the geometry of the emitting region in figure \ref{geom}

The radial distance from Sgr A$^{\ast}$ is $r^2=x^2+z^2$. For the
galactic plane (galactic latitude $b=0^o$) the coordinate $x\simeq
R_\odot\vartheta$ for small values of the galactic latitudes
$l=\vartheta$ where $R_\odot=8.5$ kpc is the distance between the GC and
Earth.

 Then the intensity of the diffuse 6.4 keV emission in the latitude
direction $\vartheta$ is
\begin{equation}
I_{6.4}(x,t)=\frac{1}{4\pi}\int\limits_{z_1}^{z_2}\epsilon_{6.4}(r,t)
dz = \eta\omega_K\sigma_0 n_H \frac{L_{fl}}{(4\pi)^2
\ln(E_x^{max}/E_x^{min})} \frac{1}{4E_0}
\int\limits_{z_1}^{z_2}\frac{dz}{x^2+z^2} \,, \label{eq_XRN}
\end{equation}
If the central source was active for the period between time momenta $\tau_2$ and $\tau_1$,
then the limits of integration are
\begin{equation}
z_1=\frac{1}{2}\left[c\tau_1-\frac{x^2}{c\tau_1}\right]
\end{equation}
and
\begin{equation}
z_2=\frac{1}{2}\left[c\tau_2-\frac{x^2}{c\tau_2}\right]
\end{equation}

Below we define $E_x^{min} = 2$ keV and $E_x^{max} = 10$ keV that correspond
to the value of $L_{fl}$ which was derived for this energy range, then
\begin{eqnarray}
I_{6.4}(x,t)=&&\frac{4.62\times 10^{13} \mbox{  ph}~\mbox{s}^{-1}~\mbox{cm}^{-1}~\mbox{sq.min}^{-1}}{x} \times \nonumber\\
&&\times\left(\frac{\eta}{2\eta_\odot}\right) \left(\frac{n_H}{\mbox{10 cm}^{-3}}\right)
\left(\frac{L_{fl}}{10^{39}\mbox{erg s}^{-1}}\right) \times\nonumber\\
&&\times\left[\arctan\left(\frac{1}{2}\left[\frac{c\tau_2}{x}-\frac{x}{c\tau_2}\right]\right)
-\arctan\left(\frac{1}{2}\left[\frac{c\tau_1}{x}-\frac{x}{c\tau_1}\right]\right)\right]\mbox{.}
\label{6_int}
\end{eqnarray}

\begin{figure}[ht]
\begin{center}
\FigureFile(100mm,100mm){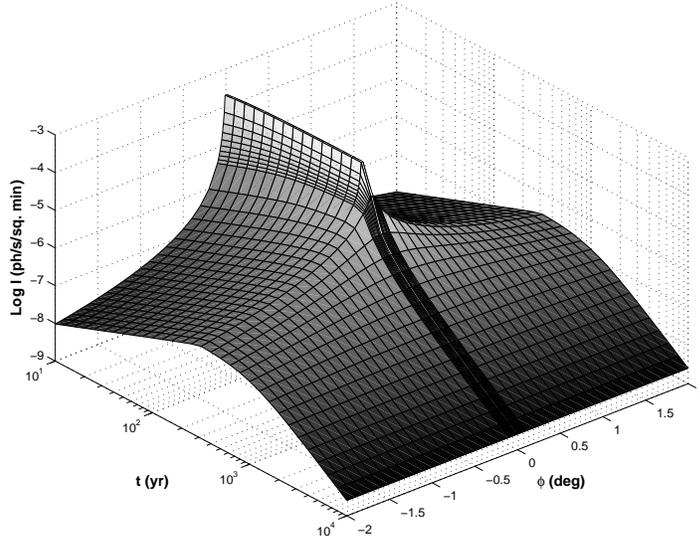}
\end{center}
\caption{Spatial and time variations of 6.4 keV line intensity in
XRN model for the SGR A$^\ast$ luminosity $L_{fl}= 1.6\times
10^{39}$erg s$^{-1}$ and the gas density $n_H=10$
cm$^{-3}$.}\label{NRX_sp}
\end{figure}

As an example we show in figure \ref{NRX_sp}  the spatial and time
variations of 6.4 keV line intensity  in the direction of the
Galactic latitude $\vartheta$ ($x=R_\odot\vartheta$) calculated
from (\ref{6_int}) when a central sources starts its activity at
$t=0$ and this activity drops to zero at $t=300$ yr.

 From this figure one can see that unlike emission
from  molecular clouds, which can be observed for short periods of
their irradiation ($\sim 10$ years), the diffuse emission  of
 6.4 keV line from the GC can be permanently seen for $\sim 10^2 - 10^3$ year even
 when a period of X-ray protons injection by Sgr A$^\ast$ is quite short.

\begin{figure}[ht]
\begin{center}
\FigureFile(100mm,100mm){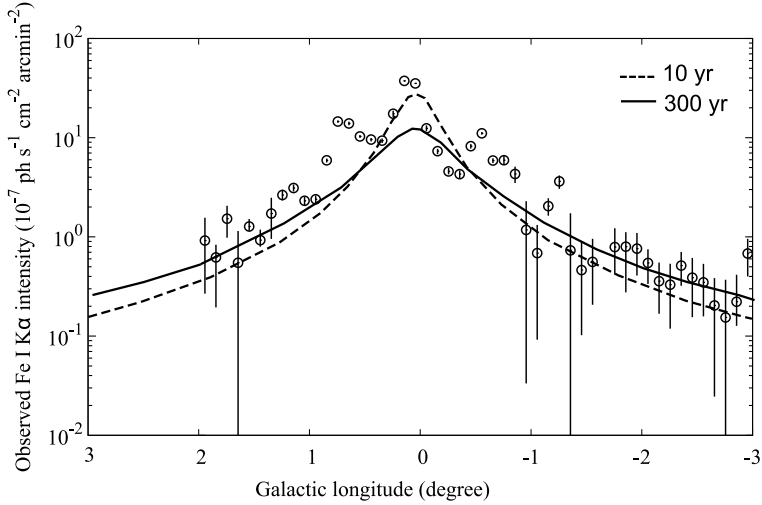}
\end{center}
\caption{Expected distribution of the diffuse 6.4 keV line along
the longitude as observed from Earth calculated in the framework
of  XRN model for the parameters: $T_1 = 10$ yr, $T = 100$ yr,
$L_{fl} = 2.9\times 10^{40}$ erg s$^{-1}$ (dashed line); and $T_1
= 300$ yr, $T = 100$ yr, $L_{fl} = 1.6\times 10^{39}$ erg s$^{-1}$
(solid line). }\label{NRX_distr}
\end{figure}

\citet{ponti} estimated the following parameters of the primary
flare: $T = 100$ yr and $T_1 = 300$ yr. The expected distribution
of the diffuse X-ray emission in the XRN model is shown in figure
\ref{NRX_distr} by the solid line. To reproduce the observed
intensity distribution of the diffuse 6.4 keV line in the GC, the
power of the central source of primary photons should be
\begin{equation}
L_{fl} = 1.6\times 10^{39} \times \left(\frac{n_H}{10\mbox{ cm}^{-3}}
\right)^{-1} \left(\frac{\eta}{2\eta_\odot}\right)^{-1} \mbox{erg s}^{-1}
\end{equation}
that is compatible with $L_{fl}$ derived for the case of Sgr B2 by
\citet{mura00}.

We notice, however, that the X-ray flare duration from Sgr
A$^\ast$ may be much shorter that estimated by \citet{ponti}.
Thus, from 6.4 keV flux variations in the direction of Sgr B2
presented in \citet{inui} the total duration of the flare   may be
about $T_1\sim 10$ years only (see also in this respect
\cite{yu11}). In figure \ref{NRX_distr} the emission distribution
for this duration of the flare is shown by the dashed line.  The
required luminosity of the flare in this case should be about
$L_{fl}\sim 2.9\times 10^{40}$ erg s$^{-1}$ that is still
compatible with the estimate of \citet{mura00} because the real
distance from Sgr A$^\ast$ to Sgr B2 may be longer than the
projection distance of 100 pc. If, however, the flare of Sgr
A$^\ast$ occurred 300 yr ago, then the required luminosity is
$L_{fl}\sim 7.8\times 10^{40}$ erg s$^{-1}$.

As it was shown in \citet{nobukawa} and \citet{dog11} the
equivalent width of the iron line provided information about the
line origin since the continuum and line X-ray emission were
generated by the same primary particles (photons or
subrelativistic charged particles). The continuum emission in XRN model is
caused by Thomson scattering of
 primary photons and it should correlate with the 6.4 keV line
emission.  The continuum emission due to the Compton echo from
molecular clouds was analysed in details by \citet{suny1}. The
intensity of photons due to the Compton scattering of primary photons
on the diffuse molecular gas can be estimated as
\begin{equation}
(dI/dE)_{c}(x)=n_H n_{ph} \int\limits_{z_1}^{z_2} F_{x}\sigma_T(\phi) dz
=0.5 n_H r_e^2 \frac{L_{fl}}{4\pi E_x^2 \ln(E_x^{max}/E_x^{min})} \int\limits_{z_1}^{z_2}\frac{(1+\cos^2\phi)dz}{x^2+z^2} \,, \label{cnt_XRN}
\end{equation}
here $\sigma_T$ is the Thomson cross-section, $r_e$ is the
classical radius of electron and $\phi$ is the scattering angle.
Taking into account that $\cos \phi = z/r$ we obtain that
\begin{equation}
(dI/dE)_{c}(x) = \frac{n_H r_e^2L_{fl}}{8\pi E_x^2 \ln(E_x^{max}/E_x^{min})} \left[ \frac{3}{x}(\arctan\frac{z_2}{x} - \arctan\frac{z_1}{x}) + \frac{z_1}{x^2+z_1^2} - \frac{z_2}{x^2+z_2^2}\right] \mbox{.}
\end{equation}

\begin{figure}[ht]
\begin{center}
\FigureFile(100mm,100mm){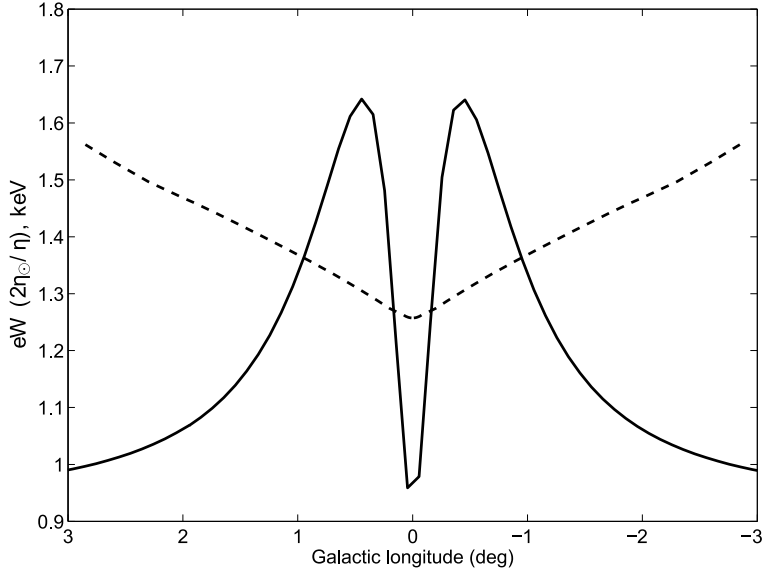}
\end{center}
\caption{Expected distribution of the equivalent width of the
diffuse 6.4 keV line along the longitude as observed from Earth
for the gas density $n_H=10$ cm$^{-3}$. Solid line is the XRN
model ($T_1 = 300$ yr, $T = 100$ yr, $L_{fl} = 1.6\times 10^{39}$
erg s$^{-1}$), dashed line is the LECRp model ($D=10^{28}$
cm$^2$s$^{-1}$, $T_c=10^4$ yr, $N_k=6\times 10^{56}$
pr).}\label{eW_distr}
\end{figure}

The distribution of the equivalent width along the Galactic
longitude expected in the framework of XRN model is shown in
figure \ref{eW_distr}. Spatial variations of {\it eW} for the XRN
model are due to the cross-section dependence on the angle
scattering.
We notice, however, it is not easy to compare this distribution of  {\it eW} with that
derived from observations because it is not easy to subtract a component of
diffuse X-ray emission produced by the Compton scattering from the
total flux of X-ray in the direction of GC: a
significant contribution of thermal emission is expected from
there. Therefore, a special procedure to subtract a Compton
component of continuum emission is necessary, as it was done e.g.
in \citet{koya09}.

\section{Spatial Distribution of the 6.4 keV Line in the LECRp Model}
Another mechanism which can generate a diffuse component of 6.4
keV emission in the GC is bombardment of the interstellar molecular gas by
subrelativistic protons whose lifetime is long enough   to fill
an extended region around the GC. As follows from \citet{dog0,dog11} these protons may be
generated by accretion processes onto the central black hole. The time-dependent
spectrum of
subrelativistic protons, $N({\bf r},E,t)$ can be calculated from
the equation (see for details \cite{dog1,dog_aa})
\begin{equation}\label{pr_state}
\frac{\partial N}{\partial t}  - \nabla D\nabla N +
\frac{\partial}{\partial E}\left( b(E) N\right) = Q(E,{\bf
r},t)\,,
\end{equation}
where  $D$ is the spatial diffusion coefficient of cosmic-ray
protons, $dE/dt \equiv b(E)$ is the rate of proton energy losses,
and $Q(E,t)$ is the rate of proton production by accretion, which
can be presented in the form
\begin{equation}
Q(E, {\bf r}, t) = \sum \limits_{k=0}Q_k(E)\delta(t -
t_k)\delta({\bf r})\,,
\end{equation}
where $t_k$ is the injection time.  The average time of star
capture in the Galaxy was taken to be $T\simeq 10^4$ years, then
$t_k=k\times T$, where $k$ is the number of a capture event.

The energy distribution of  erupted nuclei $Q_k(E)$ is taken as a
simple Gaussian
\begin{equation}\label{Qesc}
        Q_k(E)=\frac{N_k}{\sigma\sqrt{2\pi}} \exp\left[-\,\frac{(E-E_{esc})^2}{2\sigma^2}\right],
\end{equation}
where we take the width $\sigma=0.03E_{esc}$ with $E_{esc}\simeq
100$ MeV, and $N_k$ is total amount of particles ejected by each event of
stellar capture.

In the nonrelativistic case  the rate of energy losses of protons
due to Coulomb collisions can be approximated as (see e.g.
\cite{haya})
\begin{equation}
\left(\frac{dE}{dt}\right)_i\simeq \frac{4\pi n_He^4\ln\Lambda}{m_ev}\simeq \frac{a}{\sqrt{E}}\,,
\end{equation}
where $\ln\Lambda$ is the Coulomb logarithm, $v$ is the proton
velocity, $m_e$ is the electron rest mass and $a$ is a constant if
we neglect a weak dependence of the Coulomb logarithm on the
particle kinetic energy $E$. Then the solution of Eq.
(\ref{pr_state}) is
\begin{equation}
N({\bf
r},E,t)=\sum\limits_{k=0}\frac{N_k\sqrt{E}}{\sigma\sqrt{2\pi}Y_k^{1/3}}
\frac{\exp\left[-\frac{\left(E_{esc}-Y_k^{2/3}\right)^2}{2\sigma^2}-\frac{{\bf
r}^2}{4D(t-t_k)}\right]}{ \left(4\pi D(t-t_k)\right)^{3/2}}\,,
\label{sol1}
\end{equation}
where
\begin{equation}
Y_k(t,E)=\left[\frac{3a}{2}(t-t_k)+E^{3/2}\right]\,.
\end{equation}
and  $N_k$ is the total number of subrelativistic
protons emitted in each star capture event, and $T$ is the average
time of star capture.

The intensity  {$I$} of 6.4 keV line  emission in any direction ${\bf s}$ produced by
subrelativistic protons  is calculated in the same way as in \citet{dog1}
\begin{equation}
I_{6.4}({\bf{ s}})= \omega_K\eta n_H\int\limits_{ \bf
s}ds\int\limits_{E}N(E,r)v\sigma_K dE \label{is}
\end{equation}
where the integration is along the line of sight ${\bf s}$. Here
the cross-section $\sigma_K$ for subrelativistic protons was taken
from \citet{tatis}.

\begin{figure}[ht]
\begin{center}
\FigureFile(100mm,100mm){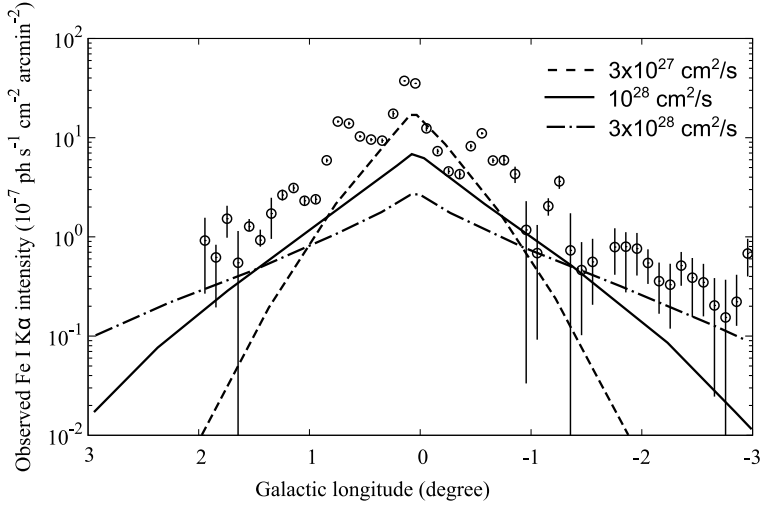}
\end{center}
\caption{Expected distribution of the diffuse 6.4 keV line along
the longitude as observed from Earth calculated in the framework
of the LECRp model for different values of the diffusion
coefficient in the GC ($T_c=10^4$yr, $N_k=6\times 10^{56}$pr).
}\label{prx_distr}
\end{figure}

The result of calculation for the LECRp model for the average gas
density $n_H=10$ cm$^{-3}$ is shown in figure \ref{prx_distr} for
different values of the diffusion coefficient in the GC.
For calculations we used the following  extreme parameters of the
proton injection: each star capture ejects $N_k=6\times 10^{56}$
subrelativistic protons, the capture frequency is $T_c=10^4$ yr
(see \cite{dog_aa}). From the figure \ref{prx_distr} one can see
that the LECRp model can also reproduce the observed diffuse 6.4
keV emission in the GC for this set of the parameters.

\begin{figure}[ht]
\begin{center}
\FigureFile(100mm,100mm){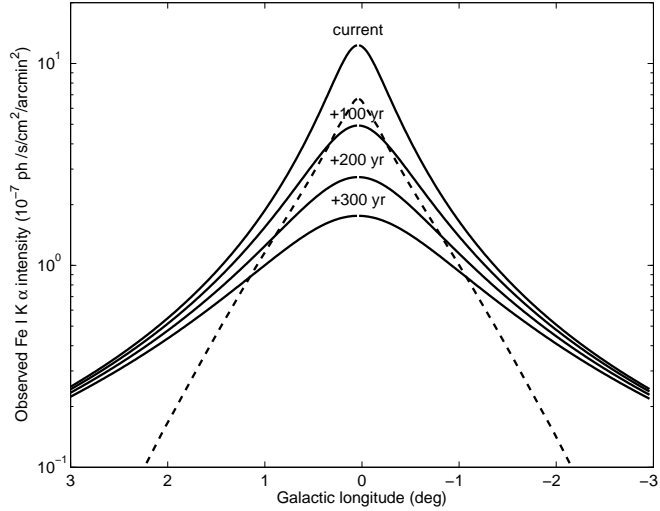}
\end{center}
\caption{Expected distribution of the diffuse 6.4 keV line along
the longitude in the future. Solid lines correspond to XRN model,
dashed line is LECRp model. Parameters of these models are the
same as in Fig. \ref{eW_distr}}\label{NRX_distr_future}
\end{figure}

In figure \ref{NRX_distr_future} we show  expected  time
variations of the 6.4 keV line emission in the XRN model and the
quasi-stationary component of 6.4 keV emission produced by
subrelativistic protons. For the both cases the gas density was
taken as $n_H=10$ cm$^{-3}$. From the figure we see that the 6.4
keV emission produced by protons may exceed that of primary XRN
photons from Sgr A$^\ast$ in 100 years from now if the parameters
of these models were chosen correctly. In this case it is highly
improbable to observe the stationary component of this line
produced by the protons from the diffuse molecular gas in the
foreseeable future.

However, we notice that if parameters of the XRN model like the
energy flux of primary photons from Sgr A$^\ast$, $L_{fl}\sim
3-10\times 10^{39}$ erg s$^{-1}$ and the delay time $T\sim
100-300$ yr and the flare duration $T_1\sim 10-300$ yr for Sgr B2
are more or less correctly estimated that makes derived values of
6.4 keV emission from the GC generated by primary photons
relatively reliable, parameters of the LECRp model are highly
uncertain. We do not know exactly which sort of stars and when was
captured by the central black hole, how many protons escape into
the GC medium, what is the diffusion coefficient there etc.

The continuum emission in LECRp model is caused by the inverse
bremsstrahlung process (see \cite{haya}). Its intensity is
\begin{equation}
(dI/dE)_{c}({\bf{ s}})= \omega_K\eta n_H \int\limits_{ \bf
s}ds\int\limits_{E}N(E,r)v\frac{d\sigma_{IB}}{dE} dE \label{ibis}
\end{equation}
where
\begin{equation}
\frac{d\sigma_{IB}}{dE} = \frac{8}{3}Z^2\frac{e^2}{\hbar c}\left( \frac{e}{mc^2}\right)^2
 \frac{mc^2}{E^\prime}\frac{1}{E_x}\ln \frac{(\sqrt{E^\prime} +
  \sqrt{E^\prime-E_x})^2}{E_x}
\end{equation}
is the cross-section of the inverse bremsstrahlung process, $E$ is
the energy of proton, $E^\prime = \frac{m}{M}E$,  $m$ is the mass
of the electron and $M$ is the mass of the proton. The
corresponding equivalent width in frame of LECRp model is shown in
figure \ref{eW_distr} by the dashed line.

\section{Discussion and Conclusion}
The diffuse  emission of the 6.4 keV line in the GC region was
recently observed with Suzaku. Only two  components can generate
ionization of the molecular gas in this extended region, namely,
hard X-ray photons or subrelativistic protons with energies about
100 MeV. Because of their long lifetime hard X-ray photons and
subrelativistic protons can propagate over large distances.

Temporal characteristics of the diffuse line emission differ from
that of compact clouds. Emission produced by photoionization in
the clouds shows temporal variations with the characteristic time
about several years that corresponds to the time in which a photon
crosses the cloud. On the other hand, the diffuse emission
generated by photionization changes with the characteristic time
 about $\lesssim 10^3$ yr. Protons in both cases generate a
 stationary flux of the line emission.

We conclude that the diffuse emission of 6.4 keV line observed at
present is  probably due to 6.4 keV vacancy production by primary
photons. This model describes nicely the observed intensity and
spatial distribution of the 6.4 keV line emission around the GC.
We notice, however, that the luminosity of Sgr A$^\ast$ required
to produce the intensity of the observed diffuse emission depends
strongly on the duration of Sgr A$^\ast$ X-ray flare. For the
delay time $T\sim 100$ yr  and the flare duration $T_1$ from 10 to
300 yr this luminosity is about $L_X\sim 10^{39} - 10^{40}$ erg
s$^{-1}$ that is compatible with the value derived by
\citet{mura00} from the observed 6.4 keV flux from the cloud Sgr
B2. If however the duration is about $T_1\sim 10$ yr and the delay
time $T\sim 300$ yr, then the required luminosity should be as
high as $\sim 10^{41}$ erg s$^{-1}$ that exceeds significantly the
estimate of \citet{mura00} derived from the Sgr B2 data.

In principle these emission can also be described in the framework
of LECRp model when the continuum and line emission is generated
by protons but in this case extreme parameters of the LECRp model
are necessary. The main problem of LECRp model is that we don't
know reliable estimates of protons injection by accretion
processes, the proton spectrum, characteristics of proton
propagation in the central region (diffusion coefficient) etc.
With all these uncertainties we can conclude that at present the
XRN model seems to be more attractive for interpretation of the
diffuse line emission in the GC than the LECRp model though we
cannot exclude that protons may contribute a significant part of
the diffuse flux.

We hope that more reliable conclusions can be obtained in the near
future. The first key results would be if observations find a
stationary component of 6.4 keV line emission from molecular
clouds. In this case the density of subrelativistic photons and a
flux of diffuse line emission generated by protons can be
estimated for the GC region.

Another very important parameter of the emission can be obtained
with the planned Astro-H mission. The point is that the width of
the 6.4 keV line produced by protons is about several tens of eV,
which is about one order of magnitude wider than the  width
expected from that generated by X-ray reflection. Future
observations by Astro-H SXS, whose energy resolution is supposed
to be only 7 eV (see \cite{taka}) will be able to measure this
parameter.

\vspace{5 mm} The authors are grateful to the unknown referee for
 careful reading of the manuscript and useful corrections. DOC
and VAD are partly supported by the NSC-RFBR Joint Research
Project RP09N04 and 09-02-92000-HHC-a. This work is supported by
Grant-in-Aids from the Ministry of Education, Culture, Sports,
Science and Technology (MEXT) of Japan, Scientific Research A, No.
18204015 (KK), and Scientific Research B, No. 20340043 (TT). This
work was also supported by the Grant-in-Aid for the Global COE
Program "The Next Generation of Physics, Spun from Universality
and Emergence" from the Ministry of Education, Culture, Sports,
Science and Technology (MEXT) of Japan.

\end{document}